%% file: CUG_2019.tex
\documentclass[10pt, conference, compsocconf]{IEEEtran}

\usepackage[utf8]{inputenc}

\DeclareFixedFont{\ttb}{T1}{txtt}{bx}{n}{7} % for bold
\DeclareFixedFont{\ttm}{T1}{txtt}{m}{n}{7}  % for normal

\newcommand\blfootnote[1]{%
  \begingroup
  \renewcommand\thefootnote{}\footnote{#1}%
  \addtocounter{footnote}{-1}%
  \endgroup
}

\usepackage{color}
\definecolor{deepblue}{rgb}{0,0,0.5}
\definecolor{deepred}{rgb}{0.6,0,0}
\definecolor{deepgreen}{rgb}{0,0.5,0}
\definecolor{brown}{rgb}{0.4,0.5,0.9}
\definecolor{purple}{rgb}{0.3,0,0.5}

\usepackage{listings}

\usepackage{amsmath}
\usepackage[hang,flushmargin]{footmisc}

\ifCLASSINFOpdf
  \usepackage[pdftex]{graphicx}
  \DeclareGraphicsExtensions{.pdf,.jpeg,.png}
\usepackage[caption=false,font=footnotesize]{subfig}
\else
\fi
\usepackage{balance}

\newcommand\pythonstyle{\lstset{
language=Python,
basicstyle=\linespread{0.9}\bfseries\ttfamily\footnotesize,
otherkeywords={self},             % Add keywords here
keywordstyle=\linespread{0.9}\bfseries\ttfamily\footnotesize\color{blue},
commentstyle=\linespread{0.9}\bfseries\ttfamily\footnotesize\color{deepgreen},
emph={MyClass,__init__},          % Custom highlighting
emphstyle=\linespread{0.9}\bfseries\ttfamily\footnotesize\color{deepred},    % Custom highlighting style
stringstyle=\color{purple},
frame=tb,                         % Any extra options here
showstringspaces=false            % 
}}

\lstnewenvironment{python}[1][]
{
\pythonstyle
\lstset{#1}
}{}

\newcommand\linuxstyle{\lstset{
language=bash,
alsoletter=-,
comment=[l][commentstyle]{//},
commentstyle=\linespread{0.9}\bfseries\ttfamily\color{deepgreen},
basicstyle=\linespread{0.9}\bfseries\ttfamily\footnotesize,
morekeywords={docker,kubectl, module, run_training,conda,pip,/bin/bash,salloc,srun,python,qsub},             % Add keywords here
keywordstyle=\linespread{0.9}\bfseries\ttfamily\color{blue},
emph=[1]{load,install,activate,pull,create,port-forward,run,expose,get,shifter,activate},
emphstyle=[1]{\linespread{0.9}\bfseries\ttfamily\color{deepred}},
emph=[2]{-N,-n,-C,-it,-v,--no-node-list,--name,-p,--image,-I,-l,--module,-t,--pty,--port,-c,--namespace},
emphstyle=[2]{\linespread{0.9}\bfseries\ttfamily\color{brown}}
,    % Custom highlighting style
stringstyle=\color{purple},
frame=tb,                         % Any extra options here
showstringspaces=false            % 
}}

\lstnewenvironment{linux}[1][]
{
\linuxstyle
\lstset{#1}
}{}

\newcommand\pythoninline[1]{{\pythonstyle\lstinline!#1!}}

\begin{document}

\title{Running Alchemist on Cray XC and CS Series Supercomputers: Dask and PySpark Interfaces, Deployment Options, and Data Transfer Times}

\input{Section_0_Authors.tex}

\maketitle

\begin{abstract}
\input{Section_0_Short_Abstract.tex}
\end{abstract}


\IEEEpeerreviewmaketitle

\blfootnote{This work appeared at CUG 2019.}

\input{Section_1_Introduction.tex}
\input{Section_2_Overview.tex}
\input{Section_3_Interfaces.tex}
\input{Section_4_RLlib.tex}
\input{Section_5_Containers.tex}

\input{Section_6_Overheads.tex}
\input{Section_7_Conclusion.tex}

% \section*{Acknowledgments}
%   This work was partially supported by NSF, DARPA, and Cray Inc.
\IEEEtriggeratref{18}
% \nocite{*}
\bibliographystyle{IEEEtran}
\balance
\bibliography{Bibliography}

\end{document}

%% file: Section_0_Authors.tex
% author names and affiliations
% use a multiple column layout for up to two different
% affiliations

\author{
  \IEEEauthorblockN{Kai Rothauge}
  \IEEEauthorblockA{
  ICSI and Dept. of Statistics\\
  UC Berkeley\\
  Berkeley, CA, USA\\
  kai.rothauge@berkeley.edu}
  \and
  \IEEEauthorblockN{Haripriya Ayyalasomayajula}
  \IEEEauthorblockA{
   Cray Inc.\\
   Seattle, WA, USA\\
   payyalasom@cray.com}
 \and
  \IEEEauthorblockN{Kristyn J. Maschhoff\hspace{6em} }
  \IEEEauthorblockA{
   Cray Inc.\hspace{6em} \\
   Seattle, WA, USA\hspace{6em} \\
   kristyn@cray.com\hspace{6em} }
\and
  \IEEEauthorblockN{\hspace{14em} Michael Ringenburg}
  \IEEEauthorblockA{
   \hspace{14em} Cray Inc.\\
   \hspace{14em} Seattle, WA, USA\\
   \hspace{14em} mikeri@cray.com}
 \and
  \IEEEauthorblockN{Michael W. Mahoney\hspace{6em} }
  \IEEEauthorblockA{
  ICSI and Dept. of Statistics\hspace{6em} \\
  UC Berkeley\hspace{6em} \\
  Berkeley, CA, USA\hspace{6em} \\
  mahoneymw@berkeley.edu\hspace{6em} }
}

%% file: Section_0_Short_Abstract.tex
Alchemist is a system that allows Apache Spark to achieve better performance by interfacing with HPC libraries for large-scale distributed computations. In this paper, we highlight some recent developments in Alchemist that are of interest to Cray users and the scientific community in general. We discuss our experience porting Alchemist to container images and deploying it on Cray XC (using Shifter) and CS (using Singularity) series supercomputers and on a local Kubernetes cluster.

Newly developed interfaces for Python, Dask and PySpark enable the use of Alchemist with additional data analysis frameworks. We also briefly discuss the combination of Alchemist with RLlib, an increasingly popular library for reinforcement learning, and consider the benefits of leveraging HPC simulations in reinforcement learning. Finally, since data transfer between the client applications and Alchemist are the main overhead Alchemist encounters, we give a qualitative assessment of these transfer times with respect to different~factors. 

%% file: Section_1_Introduction.tex
\section{Introduction}
\label{sec:introduction}

Alchemist~\cite{Gittens2018b, Gittens2018a, alchemist2019} is a system that allows Apache Spark~\cite{Zaharia2016} to achieve better performance by interfacing with high-performance computing (HPC) libraries for large-scale distributed computations. The motivation for the development of Alchemist was the inadequate performance of distributed linear algebra operations in Spark's linear algebra and machine learning module MLlib; see~\cite{Gittens2016}. It was found that not only are there significant overheads when performing the operations in Spark (up to more than an order of magnitude greater than the actual execution time of the distributed operation), but also these overheads in fact anti-scale, i.e., they increase faster than the execution time of the operation as the data sets increase in size.

Alchemist was designed to alleviate this problem by allowing users to easily interface with existing or custom HPC libraries. Efficiently implemented MPI-based linear algebra libraries do not suffer from the anti-scaling behaviour of MLlib or from large overheads not related to the execution of the actual linear algebra operation, but they are generally difficult to use for practitioners not familiar with them or with HPC in general. Alchemist therefore combines the best of both worlds: the \emph{high productivity} of Spark, allowing users to make use of its numerous data analysis components; and the \emph{high performance} of HPC libraries that can perform large-scale distributed operations faster than Spark can.

After giving a brief overview of the Alchemist framework in Section~\ref{sec:overview}, we discuss some of Alchemist's recent developments that are of interest to Cray users and the scientific community in general. Alchemist is no longer just an HPC interface just for Spark and can, in principle, be used by any data analytics framework, given a suitable client interface, and Section~\ref{sec:interfaces} introduces new client interfaces for Python, Dask and PySpark. Alchemist can also be used in applications other than data analysis and Section~\ref{sec:rllib} briefly discusses the combination of Alchemist with a reinforcement learning framework (although a detailed case study will be the subject of future research). Section~\ref{sec:containers} describes the deployment of Alchemist on different platforms using recently developed container images. While Alchemist does not suffer from the overheads that are incurred by Spark, some overheads are encountered when transmitting the data sets from the client application to Alchemist; Section~\ref{sec:overheads} tries to quantify these transfer times by taking various factors into account, namely matrix layouts, message buffer sizes, and variability in the network communication times due to varying network loads.

%% file: Section_2_Overview.tex
\section{Overview of Alchemist}
\label{sec:overview}

\begin{figure}
\centering
  \includegraphics[width=0.95\linewidth]{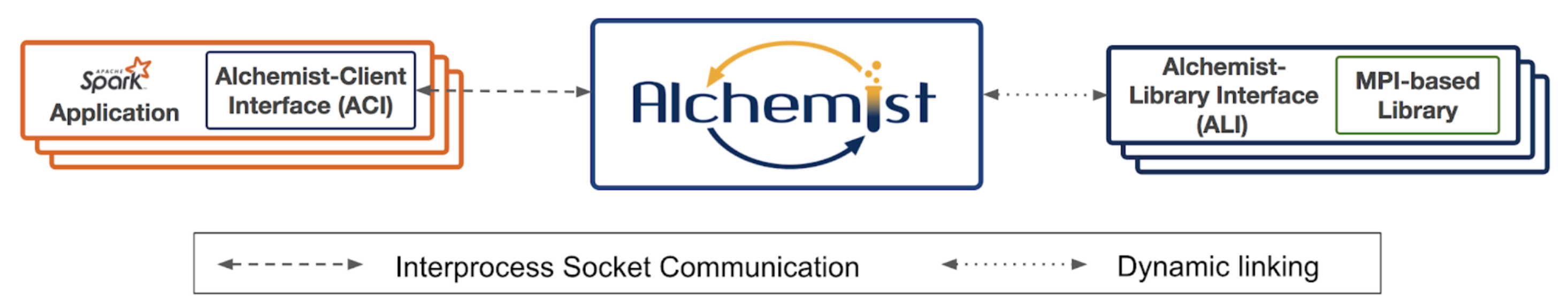}
  \caption{Overview of the basic Alchemist framework}
  \label{fig:alchemist_architecture}
\end{figure}

This section gives a brief overview of Alchemist, see~\cite{Gittens2018b, Gittens2018a} for a more extensive discussion. The basic framework of Alchemist is given in Figure~\ref{fig:alchemist_architecture}: a client application (which is a Spark application in the figure) connects to Alchemist using a suitable Alchemist-Client Interface (ACI). All communication between the client application and Alchemist occurs through the ACI. The client interface requests a number of workers from Alchemist and each of its executors connects to each of the Alchemist workers. The client interface can specify which HPC libraries it wishes to use, and these libraries are loaded by the connected Alchemist workers dynamically. Each HPC library requires a corresponding Alchemist-Library Interface (ALI) that imports the HPC library and provides wrapper functions for every function in the HPC library that is of interest to the user. It also provides a standard interface for Alchemist and calls the desired function(s) in the HPC library in the required format.

Communication between the client interface and Alchemist is primarily between the client driver process and the Alchemist driver process. If distributed data sets need to be transferred between the client interface and Alchemist, then this is done between the client workers and the Alchemist workers, where each client worker sends its portion of the data to the connected Alchemist workers. These data sets will be in the form distributed matrices that require some method of storing them, and to this end Alchemist makes use of the Elemental~\cite{elemental2017} library. Elemental is an MPI-based library that provides a convenient interface for storing distributed matrices (called \texttt{DistMatrices}), although using Elemental comes at the cost of requiring that the HPC libraries use Elemental as well so that they can access the data in the \texttt{DistMatrices}. Alchemist will also provide support for ScaLAPACK in a future version.

\begin{figure}
\centering
  \includegraphics[width=0.95\linewidth]{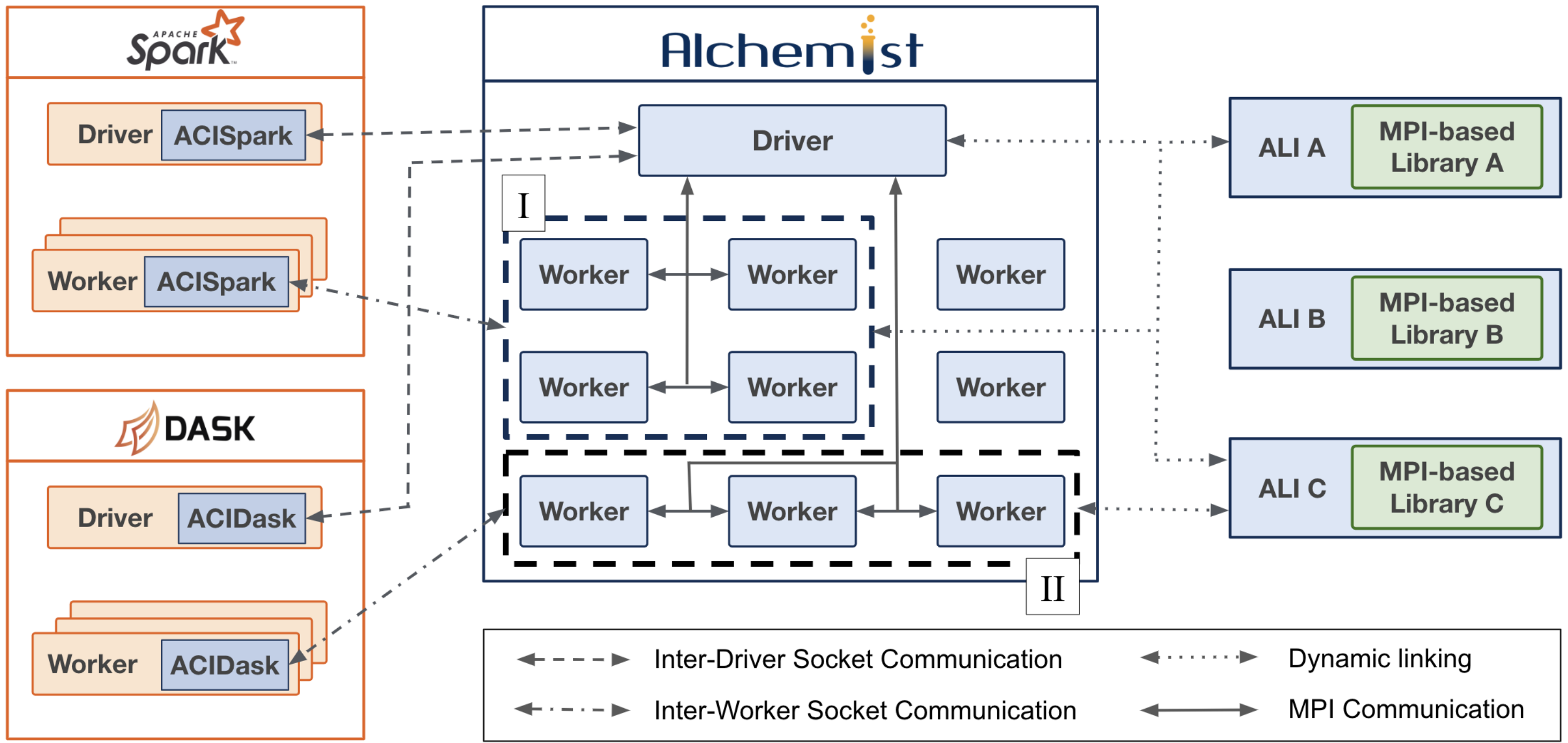}
  \caption{A representative illustration of a sample Alchemist use case. Alchemist is is running on ten nodes, with one driver node and nine worker nodes. A Spark application connects to Alchemist and requests $4$ workers, which Alchemist provides by creating a group of workers to which the Spark application can connect. The Spark application wishes to use functions in Libraries A and C, so the Alchemist workers allocated to the Spark application load these libraries dynamically. Distributed data sets are transferred between the Spark and Alchemist workers. At the same time, a Dask application connects to Alchemist and requests $3$ workers, which Alchemist provides, as well as access to the requested Library C. }
  \label{fig:alchemist_workflow}
\end{figure}

Figure~\ref{fig:alchemist_workflow} shows a schematic representation of a sample Alchemist use case, see the caption for details.

%% file: Section_3_Interfaces.tex
\section{Python, Dask, and PySpark Interfaces}
\label{sec:interfaces}

As mentioned above, Alchemist was originally written as an interface between Scala-based Apache Spark and MPI-based HPC libraries. However, recent extensions have allowed client interfaces for other languages and data analysis frameworks to be easily developed. In particular, a Python~\cite{alchemist2019} interface has been written.  It serves as a basis for client interfaces for Dask~\cite{dask}, a popular library that supports parallel computing in Python, as well as PySpark.%~\cite{pyspark}

The ability to use Alchemist from these additional frameworks enables more users to easily connect to HPC libraries. We will describe each of these interfaces in turn.

\subsection{ACIPython: Alchemist-Client Interface for Python}
\label{subsec:acipython}

\begin{figure*}
\centering
  \includegraphics[width=0.8\linewidth]{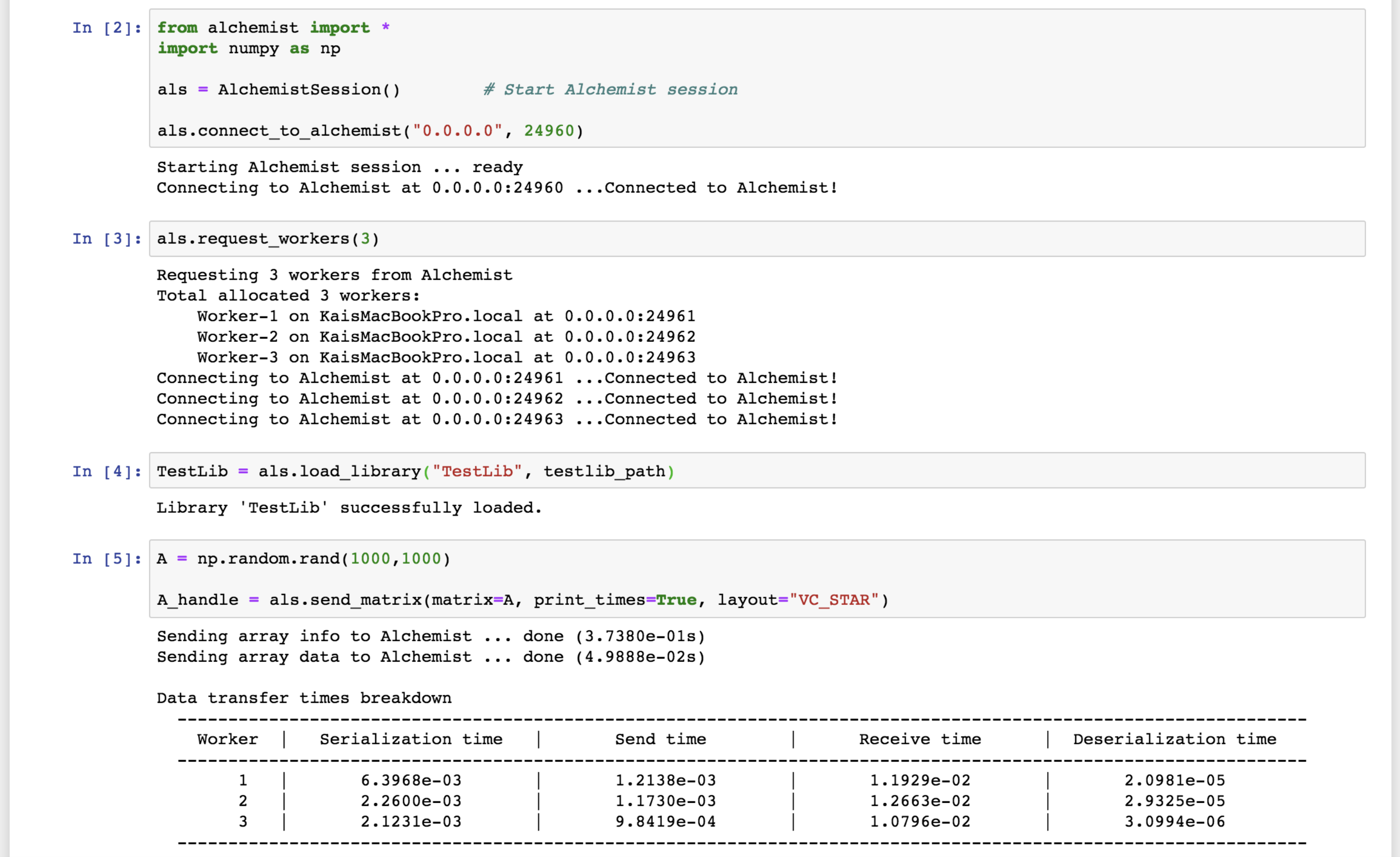} \\
  \includegraphics[width=0.8\linewidth]{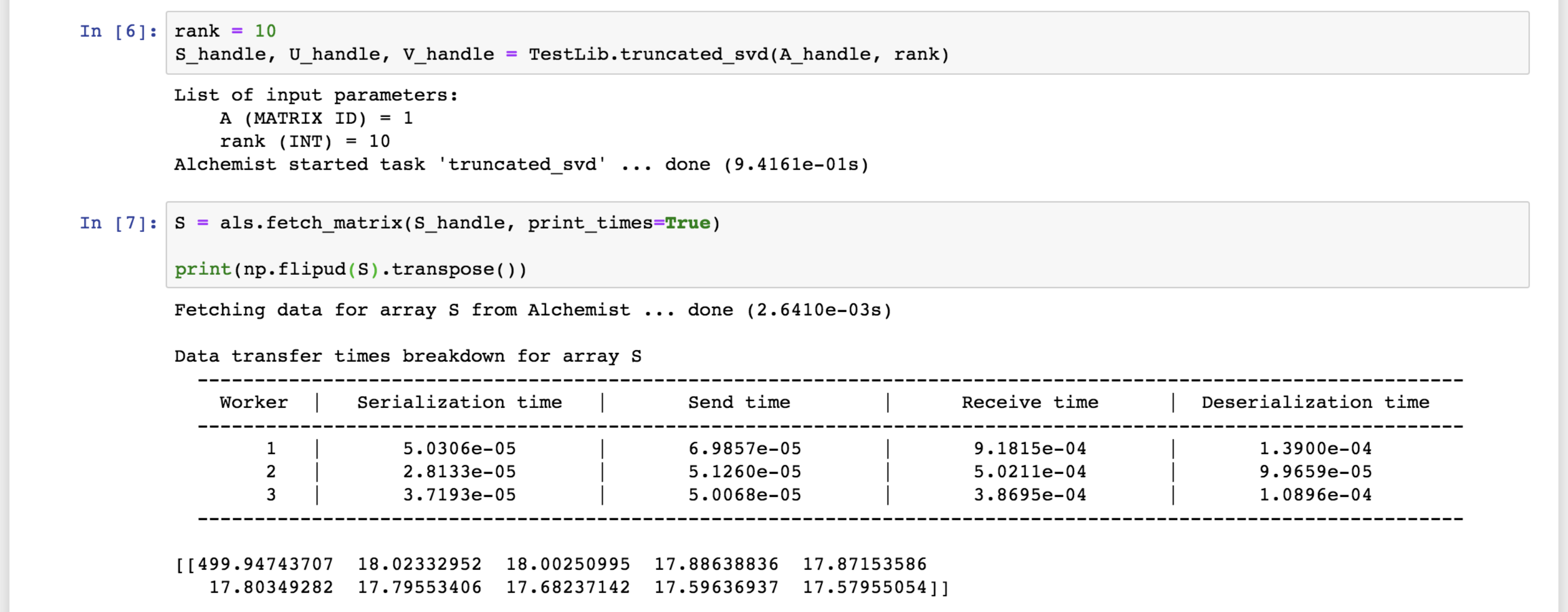}
  \caption{Screenshot of a Jupyter notebook in which Alchemist is called on a laptop using ACIPython. In this simple example the Python application connects to Alchemist, requests access to 3 workers, and loads the test library TestLib (a simple MPI-based library that provides a small set of test functions, including the truncated SVD). A randomly generated NumPy array of size $1,000\times 1,000$ is sent to the Alchemist workers, which then perform the rank-10 decomposition of it. Alchemist returns handles to each of the output matrices: the 10 left singular vectors in $U$, the 10 right singular vectors in $V$, and the singular values on the diagonal of $S$. In this case we are interested only in the singular values and we use \texttt{fetch\_matrix} to receive the entries of $S$ from Alchemist. Note that the Alchemist API may change in future versions. }
  \label{fig:acipython_notebook}
\end{figure*}

Python has established itself as the most popular language for data analysis and machine learning tasks, therefore substantial effort has been spent on the development of a client interface for Python. This allows Python users to connect to Alchemist and make use of existing HPC libraries for their data analysis and machine learning needs. We note that while there already are Python bindings for MPI (for instance the MPI4Py library~\cite{mpi4py}) that allow a Python program to exploit multiple processors, our purpose is different in that we allow users to easily connect to existing or custom HPC libraries, in particular when Alchemist is running remotely. The Python interface does not require the installation of any additional packages aside from ACIPython, and it does not require the installation of MPI, Alchemist or any of its dependencies if connecting to Alchemist remotely or when running it from inside a container.

The design of ACIPython resembles that of the Spark interface. As described above, the user connects to Alchemist via the Python interface and requests a certain number of workers. Communication is primarily with the Alchemist driver, but large matrices (or other large data sets) are sent directly to the Alchemist workers. 

An important difference is that the Python interface assumes that the underlying application is running on a single process, so that all data that is sent to Alchemist is small enough to fit in the memory of the machine that the Python application is running on. This means that at this point Python applications running on multiple processors, for instance using MPI4Py, are not yet supported, although see the ACIDask and ACIPySpark interfaces described below. The reader may question the usefulness of using Alchemist with data that is small enough to fit on a single machine, but there are several scenarios that come to mind:
\begin{itemize}
\item If the data can be loaded from a file that is accessible to Alchemist, it can be loaded by Alchemist directly (as long as it has a sufficient number of worker nodes allocated) and there is no need for the client application to load the data and transfer it.
\item Data sets that are too large to fit in the memory of a single machine can be transmitted in chunks. 
\item Intermediate stages of some computations may generate a large amount of data that will have to be stored as distributed matrices, but the input and output data sets may be significantly smaller and fit easily inside the memory of a single machine. 
\end{itemize}
%These are some of the circumstances under which Alchemist and its Python interface make it easy to access HPC libraries from a Python application, even if the Python application itself is not designed for parallel computations.
The Python interface also serves as the basis for Alchemist interfaces that do run on multiple processes, for instance the Dask and PySpark interfaces described below.

ACIPython assumes that all data sets of interest can be represented by, or converted to, NumPy arrays. The data in the array, or a subset of it, is then serialized and sent to each of the connected Alchemist workers sequentially, where they are stored in an Elemental \texttt{DistMatrix}. Each Alchemist worker receives a different chunk of the data. For instance, when transferring a $10,000 \times 10,000$ array to $10$ Alchemist workers using a row-major layout (see Section~\ref{sec:overheads}), each of the workers will receive every $10$th row of the array. Transmitting data from Alchemist back to the Python application is similarly straightforward.  In this case, the data, or a subset of it, in an Elemental \texttt{DistMatrix} is transmitted from Alchemist to the client application, where it is deserialized and stored in a NumPy array.

See the screenshot of the Jupyter notebook shown in Figure~\ref{fig:acipython_notebook} for an illustration of the use of ACIPython. 

Some users may find it useful to set up a Conda environment that packages ACIPython's dependencies:
\begin{linux}
conda create --name alchemist-python-env python=3.6
conda activate alchemist-python-env
pip install h5py
pip install pandas
\end{linux}
This environment should be started before starting the Python client application that imports ACIPython. In Section~\ref{subsec:urika} we will use this Conda environment when connecting to Alchemist on Cori.

\subsection{ACIDask: Alchemist-Client Interface for Dask}

Dask is a popular scalable data analytics platform for Python that is designed to integrate with existing applications. It provides data structures such as arrays and dataframes for storing data in larger-than-memory or distributed environments, and these parallel collections run on top of dynamic task schedulers that are optimized for computation.

ACIDask provides a convenient interface, built on top of ACIPython, for connecting Dask applications to HPC libraries using Alchemist. Our primary interest is in transmitting data stored in a Dask array to Alchemist, where it is then accessible to HPC libraries. Dask arrays are used in fields like atmospheric and oceanographic science, genomics, numerical algorithms for optimization or statistics, large scale imaging, and more; and all of these applications can potentially benefit from access to general-purpose or domain-specific HPC libraries.

Dask arrays are actually a collection of many smaller arrays, referred to as \textit{chunks} or \textit{blocks}, that may be NumPy arrays or functions that produce arrays.  If they are actual arrays, they may be stored on disk or on other machines. These arrays are arranged into a grid, and the Dask array coordinates their interaction with each other or other Dask~arrays. 

The approach taken by ACIDask is to work with the individual chunks that compose the Dask array and send them to an Elemental \texttt{DistMatrix}. Each Dask array \texttt{A} has a unique name that can be accessed using \texttt{A.name}, and every chunk in the array is referred to by the tuple \texttt{(A.name, i, j)}, with \texttt{i}, \texttt{j} being the indices of the block ranging from $0$ to the number of blocks in that dimension\footnote{Dask actually accepts up to three indices \texttt{i}, \texttt{j}, \texttt{k}, and can therefore store 3-dimensional arrays, not just matrices. Since Elemental does not support higher-dimensional arrays, we restrict ourselves to Dask arrays representing vectors or matrices.}. The $(i, j)$th chunk can be accessed by the code shown in 
the following block:

\begin{python}
# Extract the (i,j)-th chunk from a Dask Array A
def get_chunk(A, i, j):
    layers = A.dask.layers[x.name]
    a = layers[(A.name, i, j)]
    
    # Chunks are functions that produce 
    # NumPy arrays
    return a[0](*a[1])      
    # OR
    # Chunks are actual NumPy arrays
    return a[0](layers[a[1]], a[2])
\end{python}
In either case, what gets returned is a NumPy array containing the data of the $(i, j)$th chunk, which ACIDask then sends from the Dask process storing the chunk to Alchemist.

If the function in the HPC library returns a distributed matrix, Alchemist sends the dimensions of the matrix back to ACIDask, which then builds a Dask array large enough to store the data. Each Dask process then requests the data corresponding to its chunk from Alchemist and inserts it into the Dask array.

Support for Dask dataframes and other constructs may be introduced at a later date.

\subsection{ACIPySpark: Alchemist-Client Interface for PySpark}

Given that the original purpose of Alchemist was to accelerate and extend the functionality of Apache Spark when working with large, distributed data sets, it is only natural to extend the Python interface to support PySpark, the Python API for Spark, built using the Py4J library that is integrated within PySpark and allows Python to dynamically interface with JVM objects. Python generally offers improved readability of code and ease of use and maintenance compared to Scala, and PySpark has therefore become a popular interface for working with Spark's various features and libraries. For users wishing to use Spark with Alchemist, but reluctant to work with Scala, we recommend using PySpark with ACIPySpark.

As with ACISpark, ACIPySpark supports RDD-based distributed data structures defined in MLlib's \texttt{linalg.distributed} module. In particular, ACIPySpark supports \texttt{BlockMatrix}, \texttt{CoordinateMatrix}, \texttt{RowMatrix}, and \texttt{IndexedRowMatrix}, which represent distributively stored matrices backed by one or more RDDs derived from \texttt{DistributedMatrix}\footnote{As of Spark 2.0, Spark is moving to a Dataframe-based API in the \texttt{spark.ml} package for its linear algebra and machine learning operations. Support for DataFrames will be introduced in future versions of ACISpark and ACIPySpark.}.

ACIPySpark does not first convert local submatrices of a distributed matrix in PySpark into NumPy arrays before sending the data over to Alchemist. Instead, the data from the \texttt{DistributedMatrix} is serialized directly into the message buffer. Likewise, if the HPC library returns a distributed matrix, Alchemist sends the dimensions of the matrix back to ACIPySpark, which then builds a \texttt{DistributedMatrix} array to store it. Each PySpark process then requests the data corresponding to its local submatrix from Alchemist and inserts the deserialized entries into the \texttt{DistributedMatrix}.

% See Figure~\ref{fig:acipyspark_notebook.png} for a usage example of PySpark.

%% file: Section_4_RLlib.tex
\section{RLlib + Alchemist for Reinforcement Learning with HPC Simulations}
\label{sec:rllib}

Reinforcement learning (RL) \cite{sutton2018} is an area of machine learning that allows a (simulated) learner to learn by interacting with a simulated environment via a series of rewards that reflect how well the current set of parameters satisfies some set of criteria, with the goal being to maximize the number of accumulated rewards by the end of the training. The learner must find which actions to take to obtain the maximum number of rewards independently.  Therefore, due to its trial-and-error approach, a large number of simulations is required in order to train the learner successfully. While the computational cost of these simulations may be negligible when applying RL to small problems that are commonly used to illustrate its usefulness, it becomes a significant bottleneck when applying RL to large-scale problems in science and engineering that require appreciable computational resources.

It is therefore of interest to enable reinforcement learning packages to call HPC libraries for the simulations. There are potentially many areas in science and engineering that would benefit from this, in particular areas that traditionally require expensive HPC simulations and where some set of constraints and optimality conditions has to be met (airplane design, drug discovery, etc.). 

RLlib~\cite{RLlib} is an open-source library for RL that is based on the Ray~\cite{ray} framework. It provides a collection of RL algorithms and scalable primitives for composing new ones. It has seen a significant increase in interest recently, and a compelling use case of Alchemist's Python interface is in providing a simple interface through which the user of RLlib can call HPC libraries for the simulations. Alchemist thereby allows users to employ efficient HPC libraries for the simulations while still working with the extensive tool set and convenient interface provided by RLlib, hopefully facilitating the adoption of RL by the scientific and engineering communities.

A detailed case study will be the subject of future work. Here we simply give an overview of how one could call an HPC library through Alchemist inside a Python script given the current RLlib API. 

The first step is to create the class in which the simulation environment is defined:

\begin{python}
class HPCSimulator(gym.Env):
  # Initialize simulation environment
  def __init__(self, config):
  
    hostname    = config["hostname"]
    port        = config["port"]
    num_workers = config["num_workers"]
    lname       = config["lib_name"]
    lpath       = config["lib_path"]
    
    self.als = AlchemistSession()
    self.als.connect(hostname, port)
    self.als.request_workers(num_workers)
    self.HPClib = self.als.load_library(lname,
                                         lpath)
                                         
  # Reset simulator
  def reset(self):
    self.HPClib.reset()
    return self.HPClib.get_state()

  # Take a step in the simulation in response
  # to an action
  def step(self, action):
    self.HPClib.step(action)
    return self.HPClib.get_state(),
                     self.HPClib.get_score()
\end{python}
RLlib makes use of OpenAI Gym, a toolkit for developing and comparing RL algorithms. In the above sample listing, the \texttt{HPCSimulator} class is derived from OpenAI Gym's Environment class. An \texttt{AlchemistSession} is set up during initialization, and in this case we have opted that all pertinent settings are contained in a dictionary (which we called \texttt{config} here), although of course one could also read them from file. As before, we need to connect to Alchemist, request a certain number of workers, and get Alchemist to load the HPC library we want to use, denoted by \texttt{HPClib}. Presumably \texttt{HPClib} has an efficient simulator implemented that we want to use during our training procedure. To run with RLlib, \texttt{HPClib} needs to define \texttt{reset}, to set the simulator's state to its default configuration; \texttt{step}, to advance the simulation by one step in response to the \texttt{action}; \texttt{get\_state}, to return the simulators current state; and \texttt{get\_score}, to evaluate how well the current state does with regard to some problem-specific optimality condition.

To use the simulator with RLlib, we simply provide the class name as the environment within Tune~\cite{tune2018}, which is Ray's scalable hyperparameter search framework (a discussion of Tune lies outside the scope of this paper). For example:
\begin{python}
if __name__ == "__main__":
  ray.init()
  ModelCatalog.register_custom_model(...
                       "my_model", CustomModel)
  tune.run(
    "PPO",
    stop={"timesteps_total": 10000,},
    config={
      "env": HPCSimulator,
      # more configuration options ...
    }
  )    
\end{python}
See the documentation for Ray, RLlib and Tune for a clearer understanding of their APIs. The sample listings given here are just to give a flavor of what the combination of RLlib with Alchemist will look like, the actual implementation may~vary.

%% file: Section_5_Containers.tex
\section{Deploying Alchemist on Different Platforms using Containers}
\label{sec:containers}

Containers allow developers to bundle applications and their dependencies together into single entities, referred to as \textit{images}. 
Moving to a container-based deployment means users do not need to worry about building the applications from source and managing dependencies every time they want to run their application on a new platform.
Alchemist has been ported to a Docker image that can be deployed on its own on laptops and workstations, on Cray XC and CS series supercomputers, or on Kubernetes clusters. The image is deployed on the host machine and Alchemist then runs in a container that the client application can connect to via a suitable Alchemist-client interface. 

\subsection{The Alchemist Docker Image}
\label{subsec:docker}

\textit{Docker} is an open source container technology that has gained wide adoption. A \textit{Dockerfile} is a text-based configuration file that is used to assemble a Docker image; the Dockerfile contains commands for installing a base operating system, software components and shared libraries that are needed for the application, in this case Alchemist, to install and run within the image. It can be a tedious and time consuming process to install Alchemist and its dependencies, especially on new platforms, therefore a Docker image improves portability and drastically reduces build time, so that users can instead focus on their workflow.

The Alchemist Dockerfile uses a recent version of the Debian operating system and includes commands to install necessary compilers and other libraries, followed by commands to install the required dependencies and finally Alchemist itself. 

To deploy the Alchemist image locally, either on a laptop, personal computer or workstation, the following commands can be used:

\begin{linux}
// Pull the image
docker pull projectalchemist/alchemist:latest

// Run Alchemist using Docker
docker run -it --name alchemist 
  -p START_PORT-END_PORT:START_PORT-END_PORT
  projectalchemist/alchemist:latest /bin/bash
  -c "mpiexec -n NUM_PROCESSES \ 
          /data/Alchemist-main/target/alchemist \
          -p START_PORT [options]"
\end{linux}

\texttt{NUM\_PROCESSES}, \texttt{START\_PORT} and \texttt{END\_PORT} set the number of Alchemist processes and the range of ports that get opened so that the client application can connect to Alchemist and its workers. 
The Alchemist driver will run on the port \texttt{START\_PORT} and the worker processes will run on subsequent ports. It is important that all ports within this range are free and that \texttt{END\_PORT} - \texttt{START\_PORT} $\geq$ \texttt{NUM\_PROCESSES}. 
Various options can get passed to the Alchemist executable, most importantly the port that the Alchemist driver runs on using the \texttt{-p} flag. 
Additional options are available, see the online documentation. 
The client application will connect to the Alchemist driver on port \texttt{START\_PORT} using the appropriate interface and its worker processes will connect to the Alchemist worker processes on the remaining ports.

\subsection{Deploying the Alchemist Image on Cray XC and CS systems}
\label{subsec:urika}

The Alchemist Docker image can be used to deploy Alchemist on Cray XC and CS systems with relatively little effort. We provide a walk-through of how to do this using the existing \textit{Urika} infrastructure. An example of how to build and run a custom Alchemist image is also given.

Cray XC series supercomputers use \textit{Shifter}, developed at NERSC, to deploy Urika-XC container images that run Alchemist using the Docker image described in the previous subsection, whereas Cray CS series supercomputers use \textit{Singularity} for launching Urika-CS container images. \textit{Shifter} and \textit{Singularity} provide the flexibility to import Docker images without having Docker installed or being a superuser.

A Debian-based Docker image is used as the base Urika image from which the Urika-XC and Urika-CS images are created. This base image contains all shared content between Urika-XC and Urika-CS, and images specific to a given platform are created by adding in the additional platform-dependent content. These built Urika images are distributed to the various platforms by Cray and are not available on a public repository. 

The primary difference in the Urika-CS and Urika-XC Docker images relevant to us is the MPI implementation that is used: Urika-CS is OpenMPI-based, while Urika-XC is MPICH-based. To optimize performance when running on Cray systems, the goal is to utilize optimized system MPI libraries. On CS systems, system OpenMPI libraries (typically built with SLURM and CUDA support) are installed so that we can access the native fabric and GPU devices. The target on XC systems is MPICH-based Cray MPI. When building the image, stock versions of these MPI libraries are installed inside the image to allow building of MPI-based applications. The intent is to pick up the header files from the standard MPI versions inside the container (typically installed in \texttt{/usr/local/include}), but the MPI library path is set to a destination where the symbolic links to the Cray system specific libraries (shared objects) will exist when an image is started up.

Urika-XC and Urika-CS provide high-level run-time scripts that set up paths to optimized libraries and launch the container environments. The \textit{start\_analytics} script brings up an interactive analytics environment which includes setting up Spark or Dask clusters and Jupyter notebooks. The \textit{run\_training} script is used for running MPI-based tasks. An example of starting up the Alchemist driver and workers using the Urika scripts looks as follows:

\begin{linux}
// Load the analytics module
module load analytics

// Grab an allocation from the existing
// cluster resource manager (SLURM/PBS)
qsub -I -l nodes=NUM_PROCESSES

// Run Alchemist using the run_training script
run_training -v --no-node-list -n NUM_PROCESSES \  
          "/data/Alchemist-main/target/alchemist \
          -p START_PORT [options]"
\end{linux}

For users who want to build their own custom images, similar techniques can be used, although the details will be system- and site-specific.

\begin{figure*}
\centering
  \includegraphics[width=0.95\linewidth]{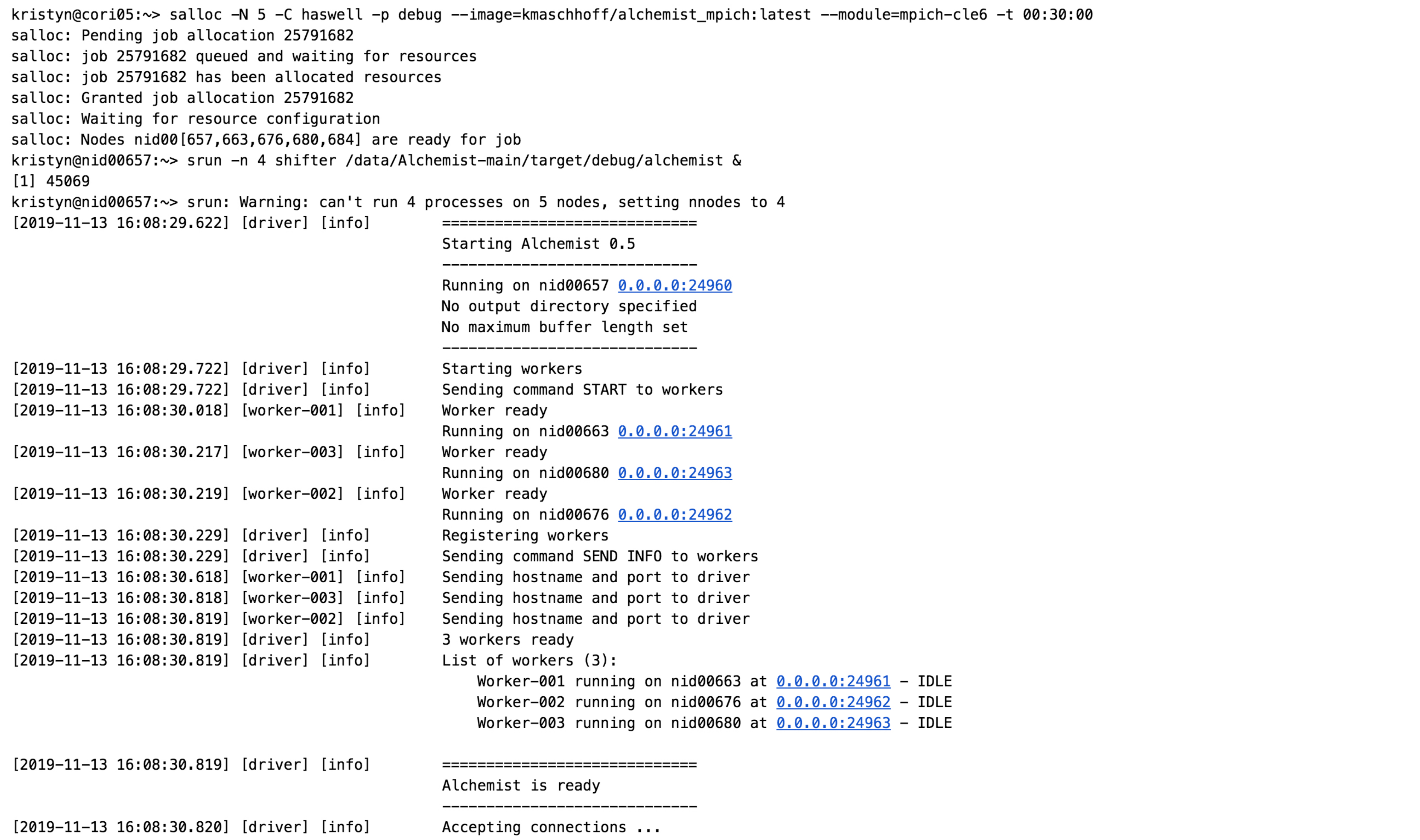} \\
  \includegraphics[width=0.95\linewidth]{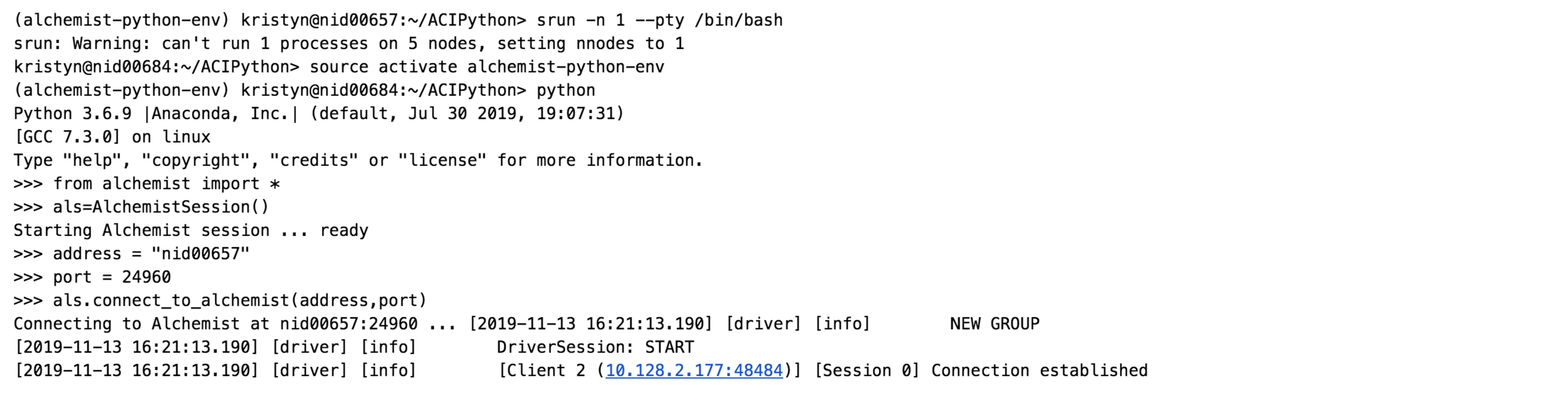} \\
  \includegraphics[width=0.95\linewidth]{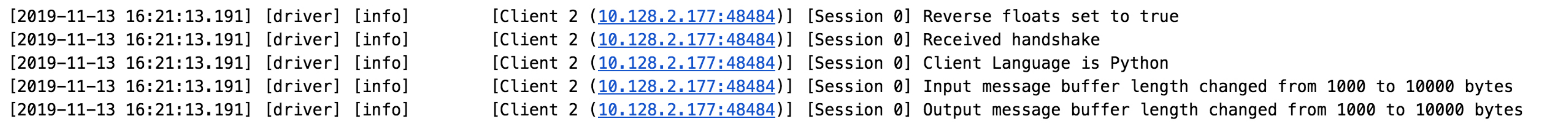} \\
  \includegraphics[width=0.95\linewidth]{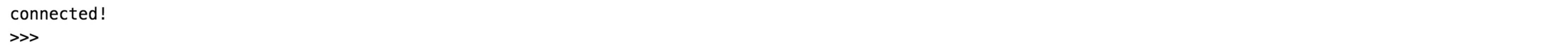}
  \caption[LoF entry]{Sample output of running Alchemist on Cori using the commands shown in the main text. After requesting five nodes using \texttt{salloc}, Alchemist is started on four of them using the \texttt{srun} command. Note that the driver is running on port 24960 on the node nid00657. This information is required by the client interface; the addresses of the nodes that the worker processes are running on will be sent to the client interface automatically and the user does not need to keep track of them. 
  
  \hspace{1em}To run a Python application with the ACIPython interface, load the packages required by ACIPython using the Conda environment \texttt{alchemist-python-env} described in Section~\ref{subsec:acipython} and start a Python shell. Import ACIPython, start the Alchemist session and connect to Alchemist by providing the address of the node that the Alchemist driver is running on and the port. 
  
  \hspace{1em}Alchemist's output can be sent to a log file so that it does not interfere with the output generated by the client application. The address and port of the node that the Alchemist driver is running on can be saved to a file of the user's choosing and loaded from file by ACIPython, see the online documentation for more details.}
  \label{fig:alchemist_cori_output}
\end{figure*}

NERSC has configured Shifter~\cite{shifternersc2019} to allow users to compile their code with standard MPICH inside their images and have the Cray libraries swapped in at run-time. A different approach can be used on standard clusters if the OpenMPI libraries built in the image are built with OFI or UCX support, such that the standard MPI implementation built inside the image can access the native network fabric.

If neither of these is an option, an approach similar to one used for Urika may be utilized. Users are responsible for making sure the necessary library shared objects are accessible from within the container and in locations where the executables expect to find them. In this case it is useful to write run-time scripts to help with the container launch, similar to the \texttt{run\_training} script used for Urika.

As a brief illustration of how one can run an Alchemist image on NERSC Cori, consider the case where the user wants to run Alchemist on four nodes on Cori and connect to it from inside a Python shell. Since Cori uses SLURM, the first step is to request an allocation using \texttt{salloc}:

\begin{linux}
// Request allocation of 5 nodes (4 for Alchemist,
// 1 for Python client application)
salloc -N 5 -C haswell -p debug \
  --image=projectalchemist/alchemist_mpich:latest
  --module=mpich-cle6 -t 00:30:00
\end{linux}
Note that we requested an additional node so that we can also run the Python shell.
The \texttt{image} flag specifies which Docker image to use, and the \texttt{module} flag specifies which Cray MPI library to swap in at run-time.

Alchemist can then be launched in the background by using \texttt{srun}:
\begin{linux}
// Run Alchemist on 4 nodes using Shifter
srun -n 4 shifter
  /data/Alchemist-main/target/alchemist &
\end{linux}

The Python shell and a Conda environment that has the necessary packages for using ACIPython, as described in Section~\ref{subsec:acipython}, is started using the commands
\begin{linux}
// Run Python client application on 1 node
srun -n 1 --pty /bin/bash
source activate alchemist-python-env
python
\end{linux}

Inside the Python shell we then import the Alchemist interface, connect to Alchemist and use it as usual. See Figure~\ref{fig:alchemist_cori_output} for a sample of the type of output one can expect when running Alchemist on Cori using the Docker image.

\subsection{Deploying the Alchemist Image on a Kubernetes Cluster}

\textit{Kubernetes} is an open source orchestration framework that supports running, scaling, and management of containers and has gained adoption both in cloud and on-premise clusters. We demonstrate running the Alchemist image on a local Kubernetes cluster, and users can follow this procedure to run Alchemist on a Kubernetes cluster deployed on a cloud~platform. 

The commands to run Alchemist on a Kubernetes cluster~are:

\begin{linux}
// Create a Kubernetes namespace for Alchemist
kubectl create namespace alchemist

// Run Alchemist using the Docker image
kubectl run -it --namespace=alchemist alchemist-k8s
  --image=projectlachemist/alchemist:latest
  --port=START_PORT <additional ports>
  -- /bin/bash -c "start_alchemist"
  
// Expose the Kubernetes deployment
kubectl expose deployment alchemist-k8s -n
  alchemist --port=START_PORT <additional ports>
  --name alchemist
  
// Get the podid
kubectl get pods -n alchemist

// Set up port forwarding
kubectl port-forward alchemist-k8s-<podid> -n
  alchemist START_PORT:START_PORT
  <additional ports>
\end{linux}

The first step is to create a Kubernetes \textit{namespace}, which is the abstraction Kubernetes uses that provides isolation to different users in a cluster. We can have multiple groups in an organization connect to different instances of the same Kubernetes cluster using different namespaces. Next, a Kubernetes deployment for Alchemist is created, which runs Alchemist in a Kubernetes \textit{pod}, the basic building block of Kubernetes.
Pods are the smallest and simplest units in the Kubernetes object model that can be created or deployed and represent a running process in the cluster.
% A pod can run any number of containers.
There are two stages involved in running Alchemist on a Kubernetes cluster: run the container on the Kubernetes cluster, then expose the ports by setting up port-forwarding to be able to connect to Alchemist from a client interface.

Note that the Kubernetes API for opening a range of ports is somewhat restricted in that we have to separately add each port that needs to be exposed. For instance, when exposing the Kubernetes deployment one would have to include the option \texttt{--port=PORT} for each port in the range (\texttt{START\_PORT}, \texttt{END\_PORT}). This would quickly become tedious when running Alchemist with a large number of processes, therefore it is recommended that the user write a script for automatically generating the Kubernetes commands with the appropriate ports exposed.

%% file: Section_6_Overheads.tex
\section{Evaluating Communication Overheads}
\label{sec:overheads}

As discussed in \cite{Gittens2018b}, the main computational overhead of Alchemist is the time it takes to transfer the data between the Spark application and Alchemist. A simple experiment to quantify these communication times for two 400GB matrices with different shapes was performed (see Tables 2 and 3 in that reference). It was observed that there is significant variability in the communication times, governed by two major factors: the number of messages sent across the network and variable network loads.

\subsection{Factors impacting communication times}

The variability of the transfer times stems mainly from variable network loads. It will generally take longer to transmit a large amount of data if the network is in heavy use, but it may also be the case that the communication between only a small number of nodes is impacted, which will still lead to a higher overall transfer time if some of the data has to be sent between these nodes. In general, we expect that a larger number of small messages will have more variability, compared to a small number of large messages, simply due to the increased likelihood that some of the messages will be delayed at some point while being transmitted across the network. Since the simulations cannot proceed until all of the data has been transferred, even one straggler can cause a higher measured transfer time.

On the other hand, it is generally more efficient for sockets to handle smaller messages, and larger messages may in fact lead to network blockages. Also, a large number of small messages sent between a large number of nodes means that more of the data is sent concurrently and one would therefore expect, under optimal conditions, smaller transfer~times.

There are several (not necessarily independent) factors that influence the number of messages sent across the network:
\begin{itemize}
    \item \textit{Amount of data}: The amount of data that needs to be sent across the network is determined by both the size of the matrix and the size of its datatypes in memory (for instance \texttt{double}s vs. \texttt{float}s).
    \item \textit{Message buffer size}: Larger buffers allow for fewer messages, but having large messages may have adverse effects, such as taking up too much memory on a core (leaving less for the actual data), and causing network blockages. 
    \item \textit{Number of Alchemist processes}: A larger number of Alchemist workers may accelerate certain computations, but it comes at the price of an increased number of messages, both between the workers during the computation and, more importantly in the context of this discussion, between Alchemist and its client interface. This is counterbalanced by the messages being shorter and more communication happening concurrently.
    \item \textit{Number of Spark partitions}: Apache Spark divides its RDDs into a number of partitions, and all tasks are then performed on these partitions in parallel, including sending the data over to Alchemist. The exact number of partitions that Spark uses depends on several factors, but generally one would expect to have at least one partition per core. With a large number of cores, this would mean that one would have a significant number of partitions that all need to connect to Alchemist concurrently to send their data, leading to a large number of small messages being sent.
    
    Since the data of a lot of these partitions is physically located on the same nodes, one would hope to be able to combine the data from different partitions destined for the same Alchemist worker before sending it across the network, but this is impossible given Spark's current API. The drawback of having each partition communicate with Alchemist directly is the number of network connections that have to be opened between the Spark application and Alchemist.
    
    Even with only a dozen nodes allocated to the Spark application and Alchemist, respectively, the number of partitions will be in the hundreds if there are a lot of cores on each node. The number of network connections will be in the thousands, and opening each of them incurs an overhead that may dominate the time it takes to send and receive the actual data if the messages are small.
    
    The rule of thumb here is that one wants to have the data in the Spark application be in the smallest number of partitions possible, i.e., each partition should hold has much data as possible, to minimize the communication overheads. In particular, one should not allocate more nodes to the Spark application than~needed.
    \item \textit{Matrix layout}: The layout used by the Elemental \texttt{DistMatrices} can have a significant impact on the performance of the HPC libraries, so it may be desirable to send the data from Spark to a \texttt{DistMatrix} that has a more favorable layout for the computations that are going to be performed on it. However, some layouts may require more messages to be sent across the network than others, for instance if a particular layout requires the local entries on one Alchemist worker to be sent from a large number of Spark partitions versus a small number.
    \item \textit{Aspect ratio of the matrix}: The aspect ratio of the matrix (its height-to-width ratio) will have an impact as well, as was found in the previous study, where sending the rows of an \texttt{IndexedRowMatrix} is more efficient and less variable if the matrix is short and wide rather than tall and thin. This is due to a smaller number of larger messages being sent, with the messages not being large enough to adversely affect communication across the network. Due to the structure of \texttt{IndexedRowMatrices} and the row-based layout used by the Elemental \texttt{DistMatrix} in that study, it also means that the partitions needed to send data to fewer Alchemist workers, leading to fewer network connections having to be opened.
\end{itemize}
We do not discuss here the time it takes to serialize and deserialize the data, but this of course affects the communication times as well. Recent improvements in Alchemist and its client interfaces have managed to decrease this overhead significantly.

Instead, here we are concerned with understanding the effect of the matrix layouts on the transmission times, but we also consider the message buffer sizes. A comprehensive study of the combined effect of all of the above factors lies outside the scope of this paper, but may be performed in future work.

\subsection{\texttt{DistMatrix} layouts}

See~\cite{elemental2017doc} for a discussion of different matrix layouts in Elemental that are possible with respect to the \textit{process grid}. The process grid is Elemental's two-dimensional arrangement of the worker processes associated with a given \texttt{DistMatrix}. For simplicity, let us assume that there are $6$ workers with IDs $1,\ldots,6$ that Elemental has arranged in a $2 \times 3$ process grid $P$:
\begin{equation*}
    P = \begin{bmatrix} 1 & 3 & 5 \\ 2 & 4 & 6 \end{bmatrix}.
\end{equation*}
There are several distribution schemes that Elemental defines, which we list here:
\begin{itemize}
    \item CIRC: Only give the data to a single process;
    \item STAR: Give the data to every process;
    \item MC: Distribute round-robin within each column of the 2D process grid (\textit{M}atrix \textit{C}olumn);
    \item MR: Distribute round-robin within each row of the 2D process grid (\textit{M}atrix \textit{R}ow);
    \item VC: Distribute round-robin within a column-major ordering of the entire 2D process grid (\textit{V}ector \textit{C}olumn);
    \item VR: Distribute round-robin within a row-major ordering of the entire 2D process grid (\textit{V}ector \textit{R}ow);
    \item MD: Distribute round-robin over a diagonal of the tiling of the 2D process grid (\textit{M}atrix \textit{D}iagonal).
\end{itemize}
The layout of a \texttt{DistMatrix} is defined by one of thirteen different legal distribution pairs \texttt{(colDist,rowDist)}. Some of these layouts allow for data to be stored redundantly (i.e., the same matrix element may be on multiple processes); since these are impractical in the large data set applications that motivate this work, we disregard these layouts. We illustrate the layouts in Elemental that do not store the data redundantly for a sample $7 \times 7$ matrix. The entries in the matrix correspond to the ID of the worker that that particular entry in the matrix is stored on.

\begin{figure*}
\centering
\begin{tabular}{l c}
  \raisebox{-0.5\height}{\small{[MC, MR]}} & \raisebox{-0.6\height}{\hspace*{-0.6em}\includegraphics[width=0.9\linewidth]{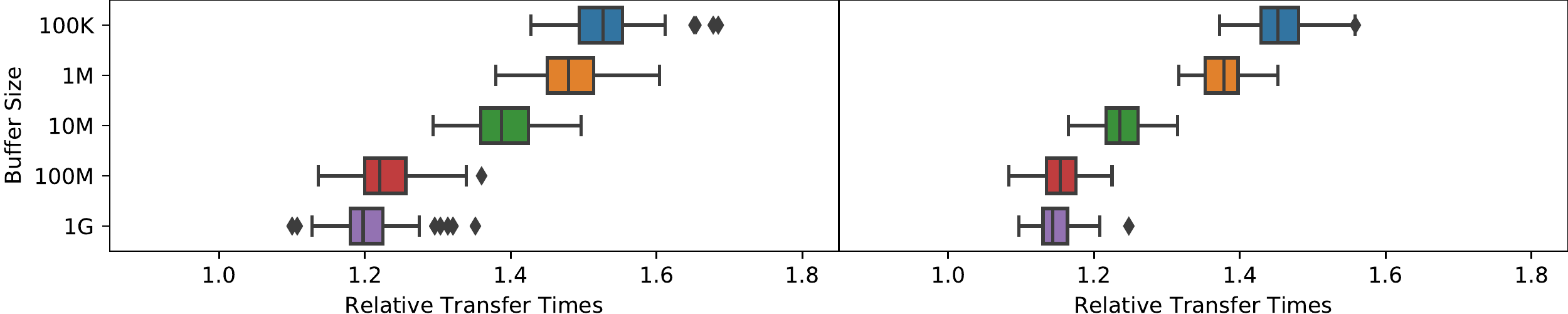}} \\[-1.8em]
  \raisebox{-0.5\height}{\small{[ * , VC]}} & \raisebox{-0.6\height}{\hspace*{-0.6em}\includegraphics[width=0.9\linewidth]{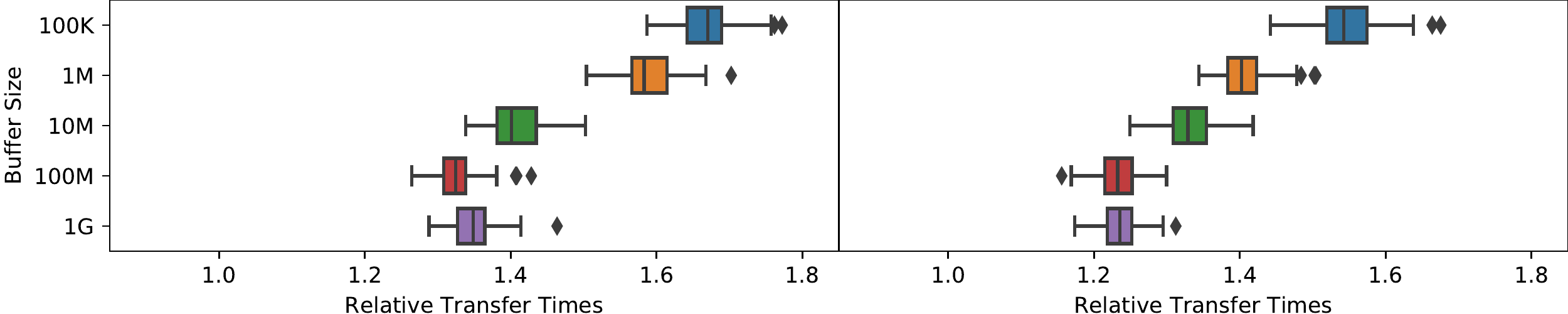}} \\[-1.8em]
  \raisebox{-0.5\height}{\small{[VC, * ]}} & \raisebox{-0.55\height}{\hspace*{-0.6em}\includegraphics[width=0.9\linewidth]{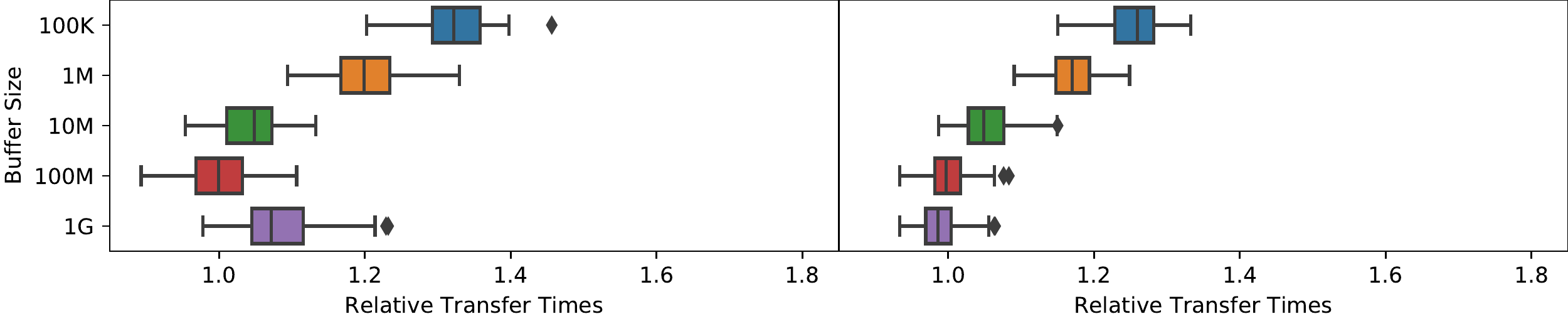}}
 \end{tabular}
  \caption{Data transfer times for various matrix layout and message buffer sizes. On the left are the transfer times for an \texttt{IndexedRowMatrix} of size $250,000 \times 200,000$, and on the right an \texttt{IndexedRowMatrix} of size $1,000,000 \times 50,000$. For each case, the transfer times are given relative to the average time it takes to transfer the data using the [VC, STAR] layout and a buffer of size 100MB. The matrix was sent from the Spark application to Alchemist $50$ times, with the transfer times represented by the box plots. Note that the transfer times decrease and become less variable as the message buffer sizes increase. }
  \label{fig:transfer_times}
\end{figure*}
\begin{itemize}
\item $[$MC, MR]: The majority of parallel routines in Elemental expect the matrices to have this layout, but it may not be the optimal layout for all purposes. Note that the process grid is tessellated with this distribution~pair.
    \begin{equation*}
    \begin{bmatrix} 1 & 3 & 5 & 1 & 3 & 5 & 1 \\ 
                    2 & 4 & 6 & 2 & 4 & 6 & 2 \\
                    1 & 3 & 5 & 1 & 3 & 5 & 1 \\ 
                    2 & 4 & 6 & 2 & 4 & 6 & 2 \\
                    1 & 3 & 5 & 1 & 3 & 5 & 1 \\ 
                    2 & 4 & 6 & 2 & 4 & 6 & 2 \\
                    1 & 3 & 5 & 1 & 3 & 5 & 1 
    \end{bmatrix}
\end{equation*}
% \item $[$MD, STAR]: Each diagonal of the tessellation of $P$ will contain the entire set of processes,
%     \begin{equation*}
%     \begin{bmatrix} 1 & 1 & 1 & 1 & 1 & 1 & 1 \\ 
%                     4 & 4 & 4 & 4 & 4 & 4 & 4 \\
%                     5 & 5 & 5 & 5 & 5 & 5 & 5 \\ 
%                     2 & 2 & 2 & 2 & 2 & 2 & 2 \\
%                     3 & 3 & 3 & 3 & 3 & 3 & 3 \\ 
%                     6 & 6 & 6 & 6 & 6 & 6 & 6 \\
%                     1 & 1 & 1 & 1 & 1 & 1 & 1 
%     \end{bmatrix}
% \end{equation*}
\item $[$MR, MC]: The transpose of the process grid is tessellated with this distribution pair.
    \begin{equation*}
    \begin{bmatrix} 1 & 2 & 1 & 2 & 1 & 2 & 1 \\ 
                    3 & 4 & 3 & 4 & 3 & 4 & 3 \\
                    5 & 6 & 5 & 6 & 5 & 6 & 5 \\ 
                    1 & 2 & 1 & 2 & 1 & 2 & 1 \\
                    3 & 4 & 3 & 4 & 3 & 4 & 3 \\ 
                    5 & 6 & 5 & 6 & 5 & 6 & 5 \\
                    1 & 2 & 1 & 2 & 1 & 2 & 1 
    \end{bmatrix}
\end{equation*}
% \item $[$STAR, MD]:
%     \begin{equation*}
%     \begin{bmatrix} 1 & 4 & 5 & 2 & 3 & 6 & 1 \\ 
%                     1 & 4 & 5 & 2 & 3 & 6 & 1 \\
%                     1 & 4 & 5 & 2 & 3 & 6 & 1 \\ 
%                     1 & 4 & 5 & 2 & 3 & 6 & 1 \\
%                     1 & 4 & 5 & 2 & 3 & 6 & 1 \\ 
%                     1 & 4 & 5 & 2 & 3 & 6 & 1 \\
%                     1 & 4 & 5 & 2 & 3 & 6 & 1 
%     \end{bmatrix}
% \end{equation*}
    \item $[$STAR, VC]:
    \begin{equation*}
    \begin{bmatrix} 1 & 2 & 3 & 4 & 5 & 6 & 1 \\ 
                    1 & 2 & 3 & 4 & 5 & 6 & 1 \\
                    1 & 2 & 3 & 4 & 5 & 6 & 1 \\ 
                    1 & 2 & 3 & 4 & 5 & 6 & 1 \\
                    1 & 2 & 3 & 4 & 5 & 6 & 1 \\ 
                    1 & 2 & 3 & 4 & 5 & 6 & 1 \\
                    1 & 2 & 3 & 4 & 5 & 6 & 1 
    \end{bmatrix}
\end{equation*}
%     \item $[$STAR, VR]:
%     \begin{equation*}
%     \begin{bmatrix} 1 & 3 & 5 & 2 & 4 & 6 & 1 \\ 
%                     1 & 3 & 5 & 2 & 4 & 6 & 1 \\
%                     1 & 3 & 5 & 2 & 4 & 6 & 1 \\ 
%                     1 & 3 & 5 & 2 & 4 & 6 & 1 \\
%                     1 & 3 & 5 & 2 & 4 & 6 & 1 \\ 
%                     1 & 3 & 5 & 2 & 4 & 6 & 1 \\
%                     1 & 3 & 5 & 2 & 4 & 6 & 1 
%     \end{bmatrix}
% \end{equation*}
    \item $[$VC, STAR]:
    \begin{equation*}
    \begin{bmatrix} 1 & 1 & 1 & 1 & 1 & 1 & 1 \\ 
                    2 & 2 & 2 & 2 & 2 & 2 & 2 \\
                    3 & 3 & 3 & 3 & 3 & 3 & 3 \\ 
                    4 & 4 & 4 & 4 & 4 & 4 & 4 \\
                    5 & 5 & 5 & 5 & 5 & 5 & 5 \\ 
                    6 & 6 & 6 & 6 & 6 & 6 & 6 \\
                    1 & 1 & 1 & 1 & 1 & 1 & 1 
    \end{bmatrix}
\end{equation*}
%     \item $[$VR, STAR]:
%     \begin{equation*}
%     \begin{bmatrix} 1 & 1 & 1 & 1 & 1 & 1 & 1 \\ 
%                     3 & 3 & 3 & 3 & 3 & 3 & 3 \\ 
%                     5 & 5 & 5 & 5 & 5 & 5 & 5 \\ 
%                     2 & 2 & 2 & 2 & 2 & 2 & 2 \\
%                     4 & 4 & 4 & 4 & 4 & 4 & 4 \\
%                     6 & 6 & 6 & 6 & 6 & 6 & 6 \\
%                     1 & 1 & 1 & 1 & 1 & 1 & 1 
%     \end{bmatrix}
% \end{equation*}
\end{itemize}
Not shown here are the layouts [VR, STAR] and [MD, STAR], which are similar to [VC, STAR] but with the rows permuted. Likewise, we do not show [STAR, VR] and [STAR, MD], since these are similar to [STAR, VC] but with the columns permuted.

Some of these layouts may not be appropriate for all cases, for instance it may not be possible to store entire rows or columns on a single process if the matrices are too wide or tall, respectively.

\subsection{Transfer time experiments}

We run our experiments on Cori~\cite{cori}, a Cray XC40 supercomputer administered by NERSC. We use its Intel Xeon ``Haswell'' processor nodes, each of which have 32 cores and 128GB of memory. Nodes on Cori communicate using the Cray-developed Aries interconnect.

For our experiment, we send a 400GB  \texttt{IndexedRowMatrix} of \texttt{double}s to an Elemental \texttt{DistMatrix} of the same dimensions. We look at the effect of the above layouts on the transfer times and also take different message buffer sizes into account. For brevity of exposition, we only consider two different matrix dimensions: $250,000 \times 200,000$ and $1,000,000 \times 50,000$. For a given layout and buffer size, the matrix is sent to Alchemist $50$ times at intervals of 30 minutes in order quantify the variability of transmission times due to network loads over a stretch of time.

On Cori, all software is managed using a modules software environment, which we use to load Spark 2.3.0. Alchemist and its dependencies are compiled from scratch and run natively on Cori, i.e., we do not use the Docker image described in Section~\ref{sec:containers}. It was found that Spark has difficulties communicating with Alchemist when running within the same job, therefore we instead run Spark and Alchemist as separate jobs concurrently, with the user connecting the Spark application to Alchemist by providing it with the hostname of the node on which the Alchemist driver is running (one should therefore start the Alchemist job before the Spark job if not running in interactive mode). For the purposes of this experiment, we run the Spark application on four nodes, and we allocate five nodes to Alchemist---one for the driver, four for the workers that will actually store the data. Since we have four Alchemist workers, the process grid will be square and there is no appreciable difference between the [MC, MR] and [MR, MC] distributions, therefore we ignore the latter distribution.

The results of the experiment are shown in Figure~\ref{fig:transfer_times}. We report the transfer times from Spark to Alchemist for the $250,000 \times 200,000$ matrix on the left, and the $1,000,000 \times 50,000$ on the right; transfer times from Alchemist to Spark are similar, so we do not show them here. Since we are interested here in the general trends shown by the transfer times, not the actual times themselves, the reported times are relative to the average transfer time of the case when Alchemist stores the distributed matrix using the [VC, STAR] layout and a 100MB buffer is used for the messages, which is Alchemist's default setting.

In general, one can conclude that it is better to have larger message buffers rather than smaller ones, but only up to a point, with 100MB seemingly a good compromise. It is generally faster to send matrices that are wider rather than narrower, although this is an artifact of \texttt{IndexedRowMatrices} storing data in rows. This also explains why sending data to Alchemist is faster if the \texttt{DistMatrix} uses a [VC, STAR] layout, since Spark is sending the data from rows to rows. In contrast, a [STAR, VC] layout requires the data in rows to be sent across columns that may be stored on different nodes by the \texttt{DistMatrix}, resulting in significantly more messages with less data and thereby increasing the overall communication times. The [MC, MR] layout is slightly more expensive than the [VC, STAR] layout since it again requires more messages to be sent, but most distributed operations will perform faster with this layout, and it is expected that it is worth the additional communication cost. This may also apply to the [STAR, VC] layout in the right context.

%% file: Section_7_Conclusion.tex
\section{Conclusion}
\label{sec:conclusion}

Several recent developments have enabled more practitioners to use Alchemist to easily access HPC libraries from data analysis frameworks such as Spark, Dask and PySpark, or from single-process Python applications. The availability of Docker and other containers enables users to get started with Alchemist quickly, and we briefly discussed the combination of Alchemist with reinforcement learning frameworks such as RLlib, which will be the subject of a more detailed future study. Alchemist's main overhead comes from the data transfer between client applications and Alchemist, and we ran some experiments to better understand the behaviour of these transfer times with respect to message buffer sizes, matrix layouts, and network variability.